# 3D Reconstruction of Coronary Vessel Trees from Biplanar X-Ray Images Using a Geometric Approach

Ethan Koland[1], Lin Xi[1], Nadeev Wijesuriya[2] and YingLiang Ma[1,2]

[1] School of Computing Sciences, University of East Anglia, UK
[2] School of Biomedical Engineering and Imaging Sciences, King's College London, UK
`e.koland@uea.ac.uk`

**Abstract.** X-ray angiography is widely used in cardiac interventions to visualize coronary vessels, assess integrity, detect stenoses, and guide treatment. In this paper, we propose a framework for reconstructing 3D vessel trees from biplanar X-ray images which are extracted from two X-ray videos captured at different C-arm angles. The proposed framework consists of three main components: image segmentation, motion phase matching, and 3D reconstruction. We developed an automatic video segmentation method for X-ray angiography to enable semantic segmentation, support both image segmentation, and motion phase matching. The goal of the motion phase matching is to identify a pair of X-ray images that correspond to a similar respiratory and cardiac motion phase; thereby reducing errors in 3D reconstruction. This is achieved by tracking a stationary object such as an injection balloon, a catheter, or a lead, within the X-ray video. The semantic segmentation approach assigns different labels to different object classes thereby enabling accurate differentiation between blood vessels, balloons, and catheters. Once a suitable image pair is selected, key anatomical landmarks (vessel branching points and endpoints) are matched between the two views using a heuristic method that minimizes reconstruction errors. This is followed by a novel geometric reconstruction algorithm to generate the 3D vessel tree. The algorithm computes the 3D vessel centrelines by determining the intersection of two 3D surfaces. Compared to traditional methods based on epipolar constraints, the proposed approach simplifies the reconstruction workflow and improves overall accuracy. We trained and validated our segmentation method on 62 X-ray angiography video sequences. On the test set, our method achieved a segmentation accuracy of 0.703. The 3D reconstruction framework was validated by measuring the reconstruction error of key anatomical landmarks, achieving a reprojection errors of 0.62 mm ± 0.38mm.

## 1    Introduction

X-ray fluoroscopy, a real-time imaging modality, is commonly used to guide cardiac interventional procedures including coronary interventions. Modern fluoroscopic systems offer high spatial and temporal resolution along with a wide field of view enabling clinicians to visualize interventional devices throughout the procedure. However, fluoroscopy provides only two-dimensional (2D) projections of inherently three-dimensional (3D) anatomical structures which limits depth perception and spatial understanding particularly during complex interventions. In addition, standard fluoroscopic images offer limited anatomical detail, as they do not clearly depict heart chambers or surrounding vasculature. To address this limitation, X-ray angiography is frequently employed to enhance anatomical visualization and assist in procedural guidance. It plays a crucial role in assessing vessel integrity, identifying stenoses (narrowing of the coronary arteries), and supporting real-time navigation. This technique involves the injection of an iodine-based contrast agent into the bloodstream which temporarily highlights the vessels. However, the enhanced visualization typically lasts only a few seconds; as the contrast agent is quickly washed out by blood flow. Therefore, clinicians often store angiographic video sequences for reference critical anatomical information throughout the procedure.

In recent years, hybrid in-procedure guidance systems have been proposed to address the limitations of X-ray fluoroscopy as a standalone guidance modality. These hybrid systems enhance minimally invasive cardiovascular procedures by integrating information from additional imaging modalities such as Magnetic Resonance Imaging (MRI) [1], Computed Tomography (CT) [2], or real-time three-dimensional transoesophageal echocardiography (3D TOE) with X-ray fluoroscopy [3][4]. In certain procedures, the availability of 3D vessel tree models within the procedural roadmap is essential. For instance, cardiac resynchronization therapy (CRT) often requires a 3D model of the coronary sinus to facilitate accurate placement of pacing leads [5]. Similarly, specific angioplasty procedures may rely on 3D models of the target vessel and its branches to ensure proper deployment of patient-specific stents.

To create detailed 3D models of blood vessel trees, we propose a 3D reconstruction framework that relies solely on two standard X-ray fluoroscopy videos acquired during contrast agent injections. These videos are recorded from two different C-arm angles using a single-plane (single C-arm) X-ray system making them a suitable dataset for 3D reconstruction. Although simultaneously captured image pairs from a biplane X-ray system are ideal for 3D reconstruction, most cardiac interventional procedures worldwide are performed using single-plane systems due to their lower cost and operational simplicity. Therefore, we developed a motion phase matching method based on semantic segmentation, which identifies both vessel boundaries and stationary objects. This allows us to select a pair of images from two X-ray video with synchronised cardiac and respiratory motion phase. We call this pair of images as "biplanar images". The tracked stationary object can be used to estimate both cardiac and respiratory motion by optimizing epipolar consistency [6]. Finally, we developed a geometric approach to reconstruct the 3D centrelines of the target vessel tree. This novel method uses Non-Uniform Rational B-Spline (NURBS) surfaces to replace traditional epipolar geometry-based techniques [7][8] thereby simplifying the reconstruction workflow and improving overall accuracy.

## 2     Method

### 2.1     Datasets

We collected a total of 62 X-ray angiography video sequences: 40 from the CADICA dataset [9] and 22 private video sequences acquired during cardiac interventional procedures using a monoplane X-ray system at St. Thomas' Hospital, London. In total, the dataset contains 2,046 individual X-ray images. These videos capture the injection of a contrast agent and its flow through the coronary arteries or the coronary sinus along the surface of the heart. Three experienced radiologists annotated the vascular regions in all 2,046 images using in-house video annotation software. For semantic video segmentation, 50 videos were used for training and validation while the remaining 12 videos were reserved for testing. For 3D reconstruction, 11 video pairs (22 videos in total) were used to evaluate reconstruction performance and calculate accuracy.

### 2.2     Semantic video segmentation

Starting from the basic U-Net architecture [10], we develop a variant version, shown in figure 1, tailored for semantic segmentation in fluoroscopy images. The proposed model has an encoder-decoder based structure with skip connections from feature encoder blocks to decoder blocks. Different from the basic U-Net, we apply the gating module after every layer of encoder to connect the decoder to generates the segmentation map $\hat{y}$. Given an intermediate feature dimension of size $f_i \in \mathbb{R}^{c_i \times h_i \times w_i}$, the output $f_i'$ of the gating module is given by $f_i' = \sigma(W \cdot f_i + b) \odot f_i$ where $i$ denotes the layer index of features, $W \in \mathbb{R}^{c_i \times c_i}$ and $b \in \mathbb{R}^{c_i}$ are learnable weight and bias parameters and $\odot$ is element-wise product along the channel dimension. Such a feature gating mechanism would suitably learn to upweight and focus on certain relevant dimensions of the feature maps that learn useful cues for the specific object segmentation, like catheter or balloon in X-ray angiography.

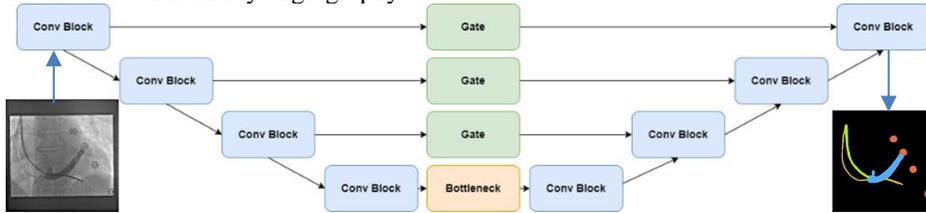

**Fig. 1.** An illustration of the proposed model. Our proposed model is consisting of three modules, an encoder, a gating module and a decoder

The feature encoder in our network is the same as the original U-Net with four convolutional blocks. Each convolutional block consists of a convolutional layer, batch normalization layer (BN), an activation function (ReLU) and a max pooling (MP) layer. The input to the first convolutional block is the input image, and the output of each block is fed to the next block. The gating module is consisting of a convolutional layer with a ReLU and follows a light-weight architecture. The feature decoder has four up convolutional blocks in total. It receives the output $f_i'$ of gating module and concatenates it before passing them to the bottleneck layer. Then, the features from the previous decoder block and features from the skip connections are up scaled using a convolutional block like the encoder, followed by transpose convolutional layers. The final segmentation map is generated by the final decoder block.

### 2.3    Motion Phase matching

Motion phase matching is based on epipolar constraints [11]. As shown in Figure 2a, if a pair of X-ray images from two video sequences are correctly matched, two key points from a stationary object will produce an intersection point or a near intersection point in 3D space. These key points could be the centres of injection balloons or the tip positions of stationary catheters or leads. The reconstructed 3D point (P) is defined as the mid-point of the shortest line segment connecting two epipolar lines (S1 and S2). Then, the reconstructed 3D point is projected back to the two image planes to obtain two 2D projected points. A 2D error is defined as the distance between the projected point and the original 2D position in one X-ray image view. The reconstruction error of the 3D point is defined as the maximum between two 2D errors. The selection of an image pair is performed by computing the reconstruction errors for all possible pairs and choosing the one with the smallest error. Figure 2b provides an example of an error matrix for all possible pairs.

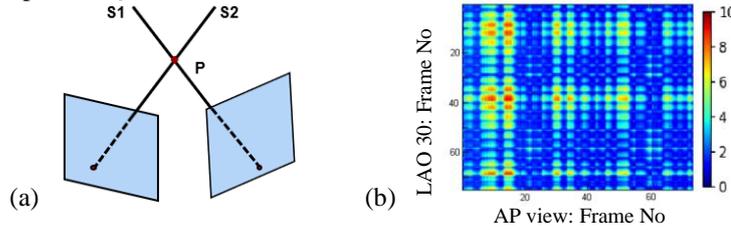

**Fig. 2.** (a) Epipolar constraint for 3D reconstruction. (b) The error matrix shows the reconstruction errors for the center of the injection balloon, with the values on the color bar expressed in millimeters (mm).

### 2.4    3D reconstruction

As illustrated in figure 3, the 3D reconstruction workflow includes the following steps: (1) Generate the centrelines and identify and match branching points. (2) Fit the centrelines using B-spline curves and divide them into separate branch curves. (3) Compute the 3D centrelines as intersection curves between two NURBS surfaces. (4) Resample the 3D centrelines to create smooth 3D vessel models.

**(1) Generate the centrelines and identify and match branching points.** Semantic video segmentation produces the binary mask of blood vessel within the selected biplanar images. A thinning algorithm [12] is applied to the binarized image and the results are one-pixel-wide skeletons. The next step is to find the branch points and end points of the skeleton. A branch point is the pixel which has more than two neighbours of skeleton pixels in connected 8-neighbours. End points only have one neighbour (figure 4(a)). Branch points and end points must be matched between biplanar images. The matched points can then be used to select corresponding pairs of centrelines for 3D reconstruction. Point-to-point matching is based on epipolar constraints. As described in section 2.3, if two points form a correct match, they will produce an intersection point or a near intersection point in 3D space. Therefore, the matched points are the one with the minimum reconstruction error.

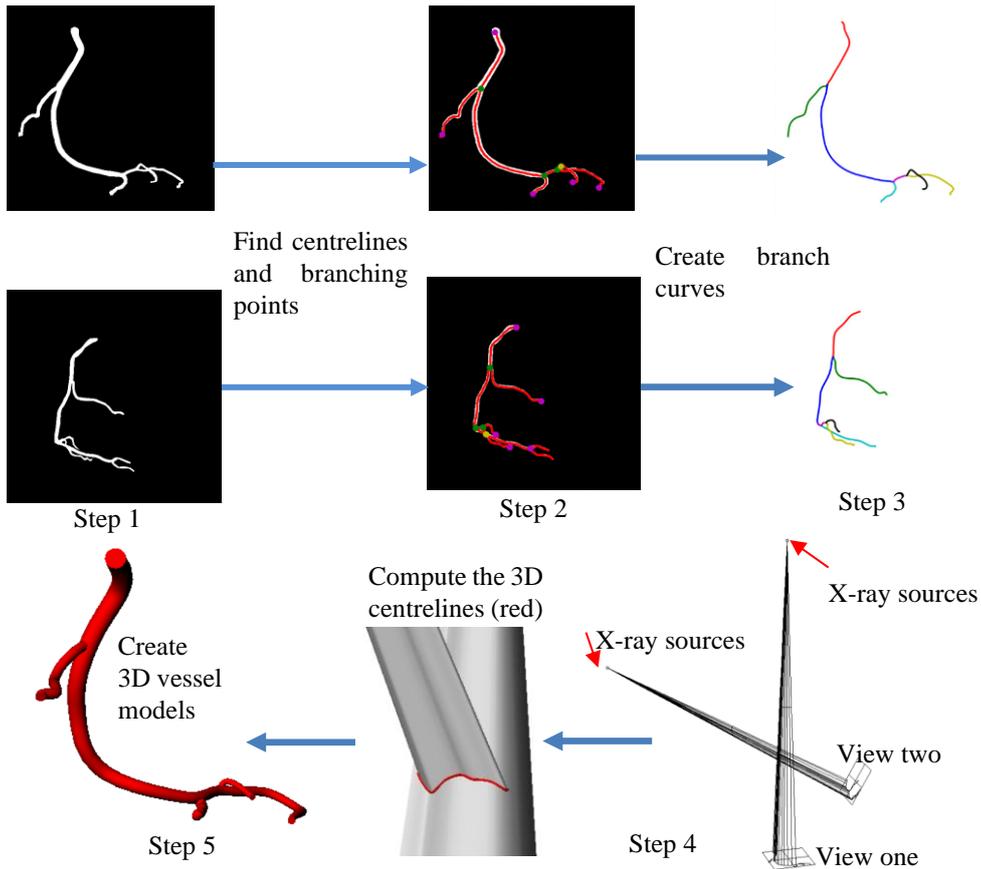

**Fig. 3.** The 3D reconstruction workflow from 2D images to 3D models.

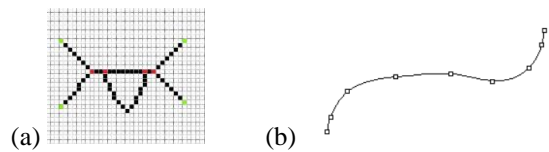

**Fig. 4.** (a) Find the branch and end points. The red points are branch points and green points are end points. (b) B-spline curve fitting for branch curves.

**(2) Create the branch curves.** After establishing matched branch points, the centrelines in the X-ray images are fitted with cubic B-spline curves (figure 4b). Cubic B-spline curves provide $C^1$ and $C^2$ continuity, enabling the creation of smooth centrelines. These curves extend either from an endpoint to a branch point or between two branch points. Step 2 in Figure 3 shows examples of such branch curves.

**(3) Compute the 3D centrelines.** As shown in Figure 5a, epipolar constraint-based reconstruction uses epipolar lines between points on the centreline curves and the X-ray radiation source. Its accuracy depends on how the points are sampled along the centrelines and on the post-processing of the intersection points between pairs of epipolar lines. To overcome these limitations, we generate the 3D centrelines of the vessel tree directly as intersection curves between two extrude NURBS surfaces. This approach eliminates many of the post-processing steps required in traditional epipolar constraint-based methods. There is no need to sort and select 3D intersection points or to reconstruct 3D curves from those points. NURBS are highly flexible and can precisely describe both simple shapes (like lines and circles) and complex organic surfaces (like car bodies or aircraft surface). A NURBS surface of degree $p$ in $u$ direction and degree $q$ in $v$ direction is defined by

$$S(u, v) = \frac{\sum_{i=0}^{n} \sum_{j=0}^{m} N_{i,p}(u) N_{j,q}(v) w_{i,j} P_{i,j}}{\sum_{i=0}^{n} \sum_{j=0}^{m} N_{i,p}(u) N_{j,q}(v) w_{i,j}}$$

Where the $\{P_{i,j}\}$ is the control net, $0 \leq u, v \leq 1$, the $\{w_{i,j}\}$ are the weights, and $N_{i,p}(u)$ and $N_{j,q}(u)$ are non-rational B-spline basis functions. In our case, the control points of the extruded surface are derived from the control points of the centreline in the X-ray images and the position of the X-ray radiation source (see Figure 5b). The calculation of intersection between two extruded surfaces can be done via the fast Newton–Raphson method [13][14].

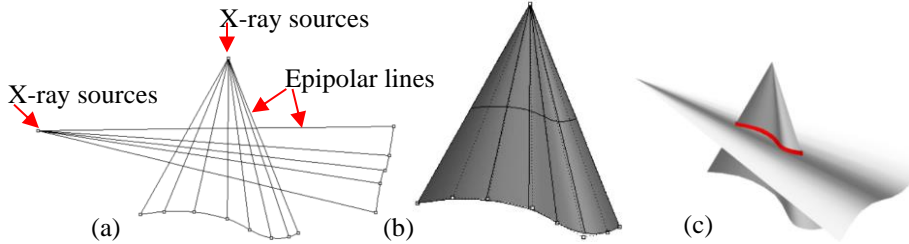

**Fig. 5.** (a) Epipolar constraint-based reconstruction (not to scale). (b) An example of extrude NURBS surface. Square points are the control points. (c) An intersection curve between two extrude NURBS surfaces. Red curve is the intersection curve.

**(4) Create smooth 3D vessel models.** The final step involves resampling the reconstructed 3D curves to further enhance their smoothness. During resampling, the NURBS curves generated from the intersection calculations are rebuilt by reducing the number of control points. This step also involves merging multiple NURBS curve segments into a single continuous curve representing the main branch of the vessel tree. To ensure smooth transitions, the tangent vector at the endpoint of one curve segment is aligned with the tangent vector of the adjoining segment, so that the entire curve achieves at least $C^1$ continuity. Finally, pipe surfaces represented as NURBS were generated using the reconstructed 3D curves. The radius of each pipe surface was determined based on the width of the segmented blood vessels in the X-ray images.

Furthermore, with some manual editing, a watertight mesh can be created, making it suitable for flow simulations or cardiac digital twin modeling.

## 3        Results

### 3.1        Semantic video segmentation

We follow the same experimental protocol for training and evaluation as adopted in prior works [15][16]. All models, including U-Net, were trained using a batch size of 8, a learning rate of $5 \times 10^{-4}$, a weight decay of $5 \times 10^{-4}$, and a maximum of 25,000 training iterations, with the Adam optimizer. All experiments were conducted using a single NVIDIA RTX 6000 GPU. Table 1 presents the mean Intersection-over-Union (mIoU) scores of our proposed method across various semantic objects, in comparison with previously reported approaches, including U-Net [10] and CMU-Net [17]. Our model achieves superior performance in semantic segmentation tasks. In addition, Figure 6 provides qualitative results illustrating the segmentation outputs of our model alongside those of U-Net and CMU-Net. These results demonstrate that our model segments objects with greater accuracy than the baseline methods.

**Table 1.** IoU score compared to related works on the X-ray angiograph dataset.

| Methods | All objects (mIoU) | Balloon (IoU) | Catheter(IoU) |
|---------|--------------------|---------------|---------------|
| U-Net   | 67.45              | 63.12         | 71.25         |
| CMU-Net | 68.17              | 65.27         | 73.29         |
| Ours    | 70.33              | 68.48         | 76.31         |

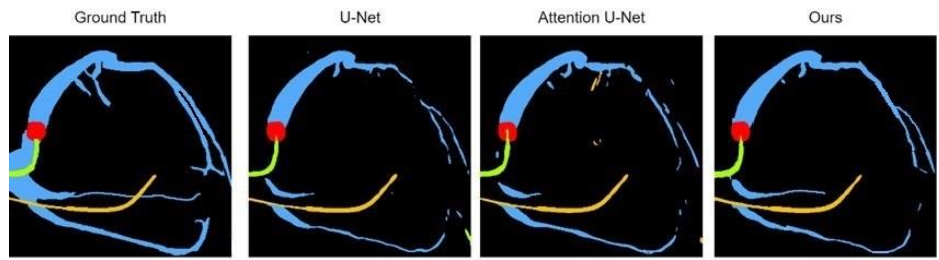

**Fig. 6.** Our method compared to U-Net and CMU-Net. Red object is the injection balloon. The yellow object is the pacing lead and the green object is the injection catheter.

### 3.2        Motion phase matching results

To test the accuracy of the motion phase matching using a stationary object, we used 22 X-ray video sequences from 11 clinical cases. Those videos were a subset of the videos we tested for semantic video segmentation. A pair of biplanar images with the minimum reconstruction error (as defined in Section 2.3) for the key point on the stationary object is selected using an error matrix. The motion phase matching error was defined as the reconstruction error of the automatically detected position of the key point on a stationary object. 0.18 mm ± 0.08 mm was achieved. Without motion phase matching, the reconstruction error of the key point in the stationary object was 1.73 mm

± 0.96 mm. The average speed of motion phase matching was 250 millisecond. This excludes the computation time of video segmentation and post-processing.

### 3.3    Whole System Validation

The performance of the proposed 3D vessel tree reconstruction framework was evaluated using datasets from 11 patients. The datasets were collected during cardiac resynchronization therapy (CRT) procedures in 4 patients and coronary interventional procedures in 7 patients. For the CRT procedures, the coronary sinus was visualized with X-ray angiography to help clinicians guide the placement of the pacing lead into one of its sub-branches. Typically, one video was acquired in the posteroanterior (PA) view, while the other was captured in either the left anterior oblique (LAO) or right anterior oblique (RAO) view. The usual X-ray C-arm angles for the oblique views are LAO 30 degrees or RAO 30 degrees. For coronary interventional procedures, the difference between C-arm angles is larger than CRT procedures. Typical angles are LAO 90 degree and LAO 30 degree. Since the goal of the proposed framework is to accurately reconstruct the geometry of the 3D blood vessel tree, the 3D centrelines are a key component. Therefore, we evaluate the accuracy of the 3D centreline reconstruction by back-projecting the 3D centrelines onto the original projection planes and calculating their 2D reprojection errors as the reconstruction error. We calculated those errors on the dataset from 11 patients and achieved an overall reconstruction error of 0.62 mm ± 0.38 mm. The entire process, including video segmentation, motion phase matching, and 3D reconstruction, takes less than 30 seconds, making it suitable for use in in-procedure guidance. Examples of reconstructed vessel trees were presented in figure 7(a)(b). Reconstruction errors for individual patient are presented in figure 7(c).

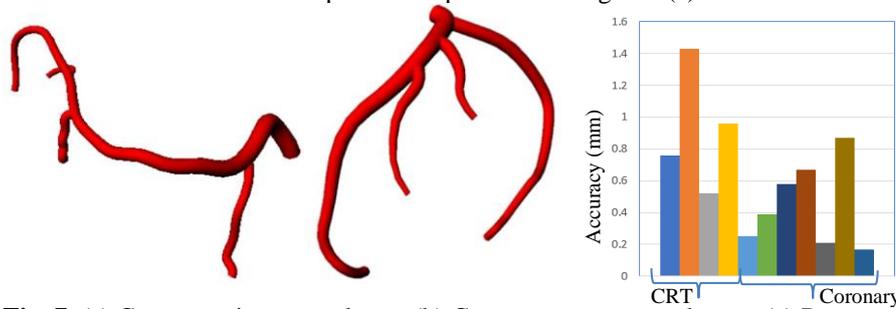

**Fig. 7.** (a) Coronary sinus vessel tree. (b) Coronary artery vessel trees. (c) Reconstruction errors for individual patients. The first four patients underwent CRT procedures, while the last seven patients underwent coronary interventional procedures.

## 4      Conclusion and Discussions

X-ray angiography plays an important role in the diagnosis and treatment of cardiovascular diseases. In this paper, we present a framework for reconstructing 3D vessel models from two X-ray fluoroscopic videos acquired at different C-arm angles. Compared to rotational X-ray angiography or 3D CT scans, using two X-ray fluoroscopic videos offers the advantage of exposing patients to less radiation during image acquisition.

The proposed framework employs a dual-purpose video segmentation method and a novel geometric approach to rapidly compute the 3D centreline of the target vessel tree. It was tested on a small dataset from 11 patients and achieved sub-millimeter accuracy based on reprojection error. A limitation of this study is that we validated the proposed framework only using reprojection error, since no 3D ground truth is available in our dataset or in the selected public dataset. A potential solution is to identify patient datasets that include both X-ray angiography and 3D CT scans.

The proposed framework has several applications in image-guided cardiac interventional procedures. It can generate a 3D vessel tree model directly from X-ray fluoroscopic video sequences, making it a feasible alternative to 3D rotational angiography, which relies on a rotating X-ray C-arm and exposes patients to significantly higher doses of radiation. In the context of hybrid guidance systems [18] for cardiac interventional procedures, our proposed framework offers unique advantages. Since the 3D model is generated directly in the X-ray image coordinate system, it eliminates the need to align the model with live fluoroscopic images. In contrast, when the 3D vessel model is extracted from CT images, an additional registration step is required to align the 3D model with 2D X-ray images for procedural guidance. Moreover, when combined with our real-time catheter tracking [19][20][21] and pacing wire tracking [22][23] methods, the proposed framework can support real-time compensation by aligning the centerlines of the 3D vessel model with the tracked catheter or pacing wire within the same blood vessel.

## 5    Acknowledgement

This work was supported by EPSRC UK (EP/X023826/1) and MRC impact fund (University of East Anglia).